\documentclass[12pt]{article}
\usepackage{amsmath}

\newcommand{\bea}{\begin{eqnarray}}
\newcommand{\eea}{\end{eqnarray}}

\newcommand{\ob}[1]{\left(#1\right)}
\newcommand{\rb}[1]{\left[#1\right]}

\title{{\small\hfill SINP/TNP/05-24, IMSc/2005/10/24}\\
\textbf{Gauge-invariant dressed fermion propagator in massless
QED$_3$}}

\author{Indrajit Mitra$^{a}$\footnote{indrajit.mitra@saha.ac.in},
Raghunath Ratabole$^{b}$\footnote{raghu@imsc.res.in}~ and
H. S. Sharatchandra$^{b}$\footnote{sharat@imsc.res.in} \\\\
$^a$ Theory Group, Saha Institute of Nuclear Physics, 1/AF Bidhan-Nagar,\\
Kolkata 700064, India\\
$^b$ The Institute of Mathematical Sciences, C.I.T. Campus, Taramani P.O.,\\
Chennai 600113, India}
\date{}
\begin{document}
\maketitle
\begin{abstract}

The infrared behaviour of the gauge-invariant dressed fermion 
propagator in massless QED$_3$ is discussed for
three choices of dressing. It is found that 
only the propagator with the isotropic (in three
Euclidean dimensions) choice of dressing
is acceptable as the physical fermion propagator. It is explained
that the negative anomalous dimension of this physical fermion
does not contradict any field-theoretical requirement.

\end{abstract}
\newpage

Massless QED$_3$ in a $1/N$ expansion \cite{appel} ($N$ being the number of fermion flavours)
has the iR (infrared) behaviour of a conformal field theory in a class of non-local gauges
\cite{mrs, mrs1}. The photon has a gauge-invariant anomalous dimension $1/2$ (scaling
dimension one), while the fermion anomalous dimension can have any value depending
linearly on the gauge parameter.
As the fermion propagator is not gauge-invariant, we may argue that
it is not relevant for physical observations. Nevertheless experiments do probe the
properties of fermions, and the calculation of the anomalous dimension of
gauge-invariant physical fermions in massless QED$_3$ has become important
in relation to the cuprate superconductors \cite{rw, ftv, 5a}. In this paper, we discuss three
choices for the gauge-invariant dressed fermion \cite{lavelle} and find that the propagator of only one
of them can be identified with the physical fermion propagator. We also explain why
the negative anomalous dimension of this physical fermion is allowed from field theory.

It has been proposed \cite{rw, ftv} that the relevant observable for the physical fermion propagator
is the `stringy ansatz' \begin{equation}
<\psi_{\alpha}(x)\exp\ob{-ie\int_{x}^{y}dz^{\mu}A^{\mu}(z)}\bar{\psi}_{\beta}(y)>\label{eq:2.1}\end{equation}
 with a `Wilson line' connecting the fermion fields. This is gauge
invariant, but depends explicitly on the path ${\cal C}$ with 
end-points $x$ and $y$. Using a straight line connecting the end-points, 
it has been claimed that for large separation this has a power law
behaviour \begin{equation}
\frac{\rlap/x-\rlap/y}{|x-y|^{3+\eta}}\label{eq:2.2}\end{equation}
 with an anomalous dimension $\eta$. 
A positive anomalous dimension could lead
to a Luttinger type behaviour and possibly explain the ARPES
data \cite{valla} . Unfortunately, an explicit calculation gives
a negative value 
$\eta=-32/3\pi^{2}N$
 in the leading order in $1/N$ \cite{khv1, rw, 8a, khv2, ye}.

However, there are  gauge-invariant observables other than the one given in  eq.\ (\ref{eq:2.1}).
Also, the choice of a straight line path between
the end points is unwarranted. For the ultraviolet behaviour, the
path is infinitesimally short, and we could expect that its choice does
not matter for the leading behaviour. But we are interested
in the iR behaviour, $|x-y|\rightarrow\infty$, where the path
${\cal C}$ becomes infinitely long and it would appear that the leading
behaviour depends on the choice.

In addition to these objections, we find a serious problem with the
choice of eq.\ (\ref{eq:2.1}). The iR behaviour is not a power
law, but has an exponential damping, depending on a cutoff. This has
to do with self-energy of the infinitely thin flux line ${\cal C}$, and is unavoidable
(See Appendix).
Thus it appears that we need to consider a more general
gauge invariant dressing for the electron than given by eq.\ (\ref{eq:2.1}).
We have to now ask which of these dressings is correct and how to
compute the correlation function including fluctuations of the flux
line.

Long back Dirac \cite{dirac} presented the gauge-invariant fermion dressed
by a photon cloud: \begin{equation}
\Psi_{\alpha}(x)=\exp\ob{-ie\int d^{3}zJ^{\mu}(x,z)A^{\mu}(z)}\psi_{\alpha}(x)\,.\label{eq:2.3}\end{equation}
Here the current $J^{\mu}(x,z)\equiv J^{\mu}(x-z)$ satisfies \begin{equation}
\partial_{z}^{\mu}J^{\mu}(x,z)=\delta^{(3)}(x-z)\label{eq:2.4}\end{equation}
 as a consequence of which, under a local gauge transformation $A^{\mu}(x)\rightarrow A^{\mu}(x)-\partial^{\mu}\theta(x)$,\begin{equation}
\exp\ob{-ie\int d^{3}zJ^{\mu}(x,z)A^{\mu}(z)}\rightarrow e^{-ie\theta(x)}\exp\ob{-ie\int d^{3}zJ^{\mu}(x,z)A^{\mu}(z)}.\label{eq:2.5}\end{equation}
 This transformation of the dressing factor compensates for the gauge transformation of $\psi(x)$. Note
that many choices of $J^{\mu}$ satisfying eq.\ (\ref{eq:2.4}) are possible.
The {\it isotropic} (in three-dimensional Euclidean space-time) {\it solution} 
is an evident
choice: \begin{eqnarray}
&&J^{\mu}(x,z)=\partial_{x}^{\mu}\phi(|x-z|)=-\partial_{z}^{\mu}\phi(|x-z|)\,,\label{eq:2.55}\\
&&\phi(|x-z|)=\frac{1}{4\pi|x-z|}\,.\label{eq:2.6}\end{eqnarray}
 As Dirac wanted a dressed operator which is local in time (though
non-local in space), he chose 
(we will refer to this choice as the {\it Dirac dressing}) \cite{dd}
\begin{eqnarray}
&&J^{3}(\vec{x},x^3;\vec{z},z^3)=0  , ~~ 
J^{i}(\vec x ,x^3; \vec z, z^3)=\delta(x^3-z^3)j^i(\vec x, \vec z)~~(i=1,2)\label{eq:2.65}\\
&&j^i(\vec x, \vec z)=\partial_{x}^{i}\phi(|\vec x-\vec z|)\label{eq:2.7}\end{eqnarray}
 In our case of $(2+1)$-dimensions, we get \begin{equation}
\phi(|\vec x-\vec z|)=-\frac{1}{2\pi}\ln|\vec{x}-\vec{z}|\label{eq:2.8}\end{equation}
 In contrast to these, choosing a {\it straight Wilson line} in eq.\ (\ref{eq:2.1}) means
squeezing this Coulomb field around the electron into a line.
If we take the line along the 3-direction of the Euclidean space-time, this choice of $J^\mu$ is given by
\bea
J^i(x,z)=0,~~J^3(x,z)=\delta^{(2)}(\vec x-\vec z)\theta(z^3-x^3)\label{eq:9a}
\eea
(this is explained after eq.\ (\ref{eq:2.10})).
It is
interesting to note that the current of eq.\ (\ref{eq:9a})  differs from the isotropic
choice of eqs.\ (\ref{eq:2.55})  and (\ref{eq:2.6}) by precisely the curl of
the vector potential used by Dirac for the magnetic
monopole \cite{god}. The infinitely squeezed flux (electric and not magnetic in the present case)
is the Dirac string resulting from the singular vector potential.

The two-point correlation function which results from eq.\ (\ref{eq:2.3}) is 
\begin{eqnarray}
{\cal S}_{\alpha\beta}(x,y) & = & <\Psi_{\alpha}(x)\bar{\Psi}_{\beta}(y)>\\
 & = & <\psi_{\alpha}(x)\exp\ob{-ie\int d^{3}z {\cal J}^{\mu}(z)A^{\mu}(z)}
 \bar{\psi}_{\beta}(y)>\label{eq:2.9} \end{eqnarray}
with \bea
{\cal J}^{\mu}(z)\equiv {\cal J}^{\mu}(x,y;z)=J^{\mu}(x,z)-J^{\mu}(y,z)\,. \label{eq:2.95}
\eea
From eq.\ (\ref{eq:2.4}), it then follows that
\begin{equation}
\partial_{z}^{\mu}{\cal J}^{\mu}(z)=\delta^{(3)}(x-z)-\delta^{(3)}(y-z)
\,.\label{eq:2.10}\end{equation}
[It may be noted that the use of eqs.\ (\ref{eq:9a}) and (\ref{eq:2.95}) in eq.\ (\ref{eq:2.9}) 
gives
\bea
<\psi_{\alpha}(x)\bar\psi_\beta(y)\exp\ob{-ie\int_{x_3}^{y_3}dz_3 A_3(z)}>\,.\label{eq:wil}\eea
Thus, as stated before, we have a straight Wilson line which lies along the 3-axis, i.e., $x$ and $y$ are separated only in the 3-direction.]

We may consider all three choices of $J^{\mu}$
mentioned above. It is revealing to express the exponent
of eq.\ (\ref{eq:2.9}) using the Fourier components $\tilde{A}^{\mu}(l)$
of $A^{\mu}(z)$. The three choices correspond to \begin{eqnarray}
 & (1) & <\psi_{\alpha}(x)\bar{\psi}_{\beta}(y)\exp\ob{-ie\int\frac{d^{3}l}{(2\pi)^{3}}(e^{il.x}-e^{il.y})\frac{il^{\mu}}{l^{2}}\tilde{A}^{\mu}(l)}>\,,\label{eq:2.11}\\
 & (2) & <\psi_{\alpha}(x)\bar{\psi}_{\beta}(y)\exp\ob{-ie\int\frac{d^{3}l}{(2\pi)^{3}}(e^{il.x}-e^{il.y})\frac{i\vec{l}}{\vec{l}^{2}}\cdot\vec{\tilde{A}}(l)}>\,,\label{eq:2.12}\\
 & (3) & <\psi_{\alpha}(x)\bar{\psi}_{\beta}(y)\exp\ob{-ie\int\frac{d^{3}l}{(2\pi)^{3}}(e^{il.x}-e^{il.y})\frac{il^{3}}{(l^{3})^{2}}\tilde{A}^{3}(l)}>\,.\label{eq:2.13}\end{eqnarray}
In the case (3), the electric flux is tightly squeezed in the $3$-direction.
The case (2) can be obtained from fluctuations of this flux line in
the space-like directions, $x_{3}=$constant and $y_{3}=$constant.
The case (1) entails fluctuations in all space-time directions.
Some details of the derivation
of eqs.\ (\ref{eq:2.11})-(\ref{eq:2.13}) are given in the Appendix.              

Eq.\ (\ref{eq:2.1}) has been calculated in the leading order by many
authors \cite{khv1, rw, 8a, khv2, ye} using diverse techniques. Calculation using the temporal gauge, which 
simplifies the contribution of the Wilson
line, yields a $\ln|x-y|$ term. This is supposed
to exponentiate to yield a power law. However there is also an iR
divergence which is evaded by deforming the contour of integration \cite{ye}.
When the calculation is performed in the (non-local) covariant gauge, a cut-off dependent
term linear in $|x-y|$ appears. This seems to preclude a power-law 
fall-off \cite{5a}. Presuming that this linear term is an artifact
of the gauge-noninvariant momentum cut-off and ignoring it, the remaining $\ln|x-y|$
dependence matches with the calculation in the temporal gauge as required
by the gauge invariance of the correlation function \cite{ye}.

In the Appendix, we calculate eq.\ (\ref{eq:2.1}) in the leading order, in the (non-local) covariant
gauge, and {\it with a gauge invariant regulator}. We find that the term
linear in $|x-y|$ persists (albeit with a $\ln\Lambda$ coefficient).
On exponentiation this gives a correlation function that does not
fall off as a power, but falls off exponentially at large distance.
In the Appendix, we also show that the Dirac dressing gives a divergent propagator
even for finite separation between the fermions. {\it There are no such problems with the
isotropic dressing.}

We now give a physical argument for the isotropic choice of dressing turning out to be the only one which is acceptable.
Considering eq.\ (\ref{eq:2.9}) in the path-integral representation, we see that the dressed fermion
propagator represents fermion propagation in the presence of a source ${\cal J}^\mu$
for the electromagnetic field $A^\mu$. In three dimensions, ${\cal J}^\mu$ is in general the
sum of a gradient and a curl. Integration by parts shows that 
the curl part of ${\cal J}^{\mu}$
couples to the field strength $F^{\mu\nu}$; so it is truly a source
for the physical degrees of the photon. Thus if ${\cal J}^\mu$ (or $J^\mu$) has
a curl part, 
we are computing the propagation of fermions in the presence of an external
physical electromagnetic source. {\it The only 
dressing that corresponds to physical fermions in the absence of external
electromagnetic sources is the one in which} $J^{\mu}$ {\it is a pure gradient}.
This singles out the choice given in eqs.\ (\ref{eq:2.55}) and (\ref{eq:2.6}),
and so Wilson line fluctuations in space-time have to
be permitted.

It has been argued by Kennedy and King \cite{kk} (see also Ref.\ \cite{ft}) that {\it the gauge-invariant
correlation function with the isotropic dressing reduces to just the usual
fermion propagator in the Landau gauge}. The reason is that the
Landau gauge corresponds to simply inserting $\partial A=0$ constraint
in the functional integral \cite{not}: \begin{equation}
\lim_{\alpha\rightarrow0}\int{\cal D}A\exp\ob{-\frac{1}{2\alpha}\partial A\cdot\partial A}\cdots=\int{\cal D}A\delta\rb{\partial A}\cdots\label{eq:2.15}\end{equation}
From eqs.\ (\ref{eq:2.95}) and (\ref{eq:2.55}), 
\begin{equation}
\int d^{3}z{\cal J}^{\mu}(x,y;z)A^{\mu}(z)=\int d^{3}z\ob{\phi(|x-z|)-\phi(|y-z|)}\partial_{z}^{\mu}A^{\mu}(z)\label{eq:2.16}\end{equation}
As only the $\partial A$ part
is coupled to the sources, the source term vanishes and the
dressing factor is unity in the Landau
gauge.

This result permits us to read off the iR behaviour of the gauge invariant
dressed fermion propagator. For this note that {\it even with our
non-local gauges}, the $\alpha\rightarrow 0$ limit
corresponds to the constraint $\partial A=0$ \cite{not}:
\begin{equation}
\lim_{\alpha\rightarrow0}\int{\cal D}A\exp\ob{-\frac{1}{2\alpha}\partial A\cdot g\cdot\partial A}\cdots\propto\int{\cal D}A\delta\rb{\partial A}\cdots\label{eq:2.17}\end{equation}
 As the usual fermion propagator has a power law behaviour with an anomalous
exponent $4(\alpha-\alpha_{0})/(\pi^{2}N)$ \cite{mrs1}, we get the iR behaviour
of the dressed fermion by simply choosing the Landau
gauge value $\alpha=0$: \bea
{\cal S}(x-y)\sim\frac{\rlap/x-\rlap/y}{|x-y|^{3-\frac{4\alpha_{0}}{\pi^{2}N}}}
\eea
implying an anomalous dimension of (see eq.\ (\ref{eq:2.2}))
$\eta=-4\alpha_0/(\pi^2 N)$. 
Note that $\alpha_{0}$ is
a series in $1/N$ with the leading order value $2/3$ \cite{xi}.
Therefore $\eta=-8/(3\pi^2 N)$ to
the leading order in $1/N$. 

It is to be noted that the anomalous dimension is negative.
We now
argue that {\it there is no contradiction with the 
fact} \cite{pos} {\it that the properties  of the spectral function
imply a non-negative anomalous dimension}.
Note that the physical correlation ${\cal S}(x,y)$ of eq.\ (\ref{eq:2.9})
does not involve local operators at only $x$ and $y$. The physical
fermions of eq.\ (\ref{eq:2.3}) are not localized in space-time. The Kallen-Lehmann
spectral representation is not valid for correlations between such non-local objects.
Of course in
the Landau gauge, the physical correlation function does reduce to
one between local operators at $x$ and $y$. But the Hilbert space of
the theory in any covariant gauge contains states of negative norm. In that case, even though the
Kallen-Lehmann spectral representation is valid, the positivity of the 
spectral function
is not guaranteed.

In this paper, we have discussed three possible choices of the gauge-invariant dressed fermion 
as candidates for the physical fermions  of massless QED$_3$. It appears from our analysis that
the physical fermions possess a negative anomalous dimension.

\section*{Acknowledgements}

We thank D. Khveshchenko, M. Lavelle and Z. Tesanovic for useful communications.
I.M. thanks IMSc, Chennai for hospitality during the course of this work.

\appendix

\section*{Appendix}

In this Appendix we present the features of the gauge-invariant fermion
propagator of eq.\ (\ref{eq:2.9}) in the leading order in $1/N$. {\it We treat
all the three dressings given in eqs.\ (\ref{eq:2.11})-(\ref{eq:2.13}) in the
same way in order to compare and contrast them}. As they are invariant,
we should get the same answer in any gauge. However, there are prescription
ambiguities when the temporal gauge and the Coulomb gauge are used. Therefore we
avoid them and use an explicitly Lorentz covariant gauge. We also
avoid gauge non-invariant cut-offs, as they may lead to spurious results.
Our regularization is by using sufficient number of higher derivatives
in the photon kinetic energy terms.
 
To find the contributions to the dressed fermion propagator to leading order,
consider the numerator of the path-integral representation for the expression
(\ref{eq:2.9}):
\bea
\int{\cal D}\psi{\cal D}\bar\psi{\cal D}A \,\,\psi_\alpha(x)\bar\psi_\beta(y)
 \exp\ob{-ie\int d^{3}z {\cal J}^{\mu}(z)A^{\mu}(z)}
 \exp\ob{e\int d^{3}w \,\bar\psi(w)\rlap/A(w)\psi(w)}
\eea
(here the free part of the QED action is taken to be  understood).
Expanding the two exponentials upto $O(e^2)$, we find {\it the following three contributions to
the leading order} after performing Wick contractions:

(A) The usual one-loop fermion self-energy is obtained by considering the $O(e^2)$
term from the QED interaction Lagrangian.

(B) The self-energy of the photon cloud is obtained by considering the $O(e^2)$
term from the dressing factor.

(C) The interaction of the photon cloud with the fermion is obtained by considering
the $O(e)$ term from each exponential.

It is easy to check that {\it the net contribution is independent of the
gauge parameter} $\alpha$. The part of the photon propagator $\Delta^{\mu\nu}(x,y)$
depending on $\alpha$ is longitudinal in both $x$ and $y$. We denote
this part as \begin{equation}
\alpha\partial_{x}^{\mu}\partial_{y}^{\nu}h(x,y)\label{eq:A.1}\end{equation}
where $h(x,y)\equiv h(x-y)=h(y-x)$.
 The contributions (A), (C) and (B) from this part are, respectively, 
\begin{equation}
e^{2}\alpha\int d^{3}z_{1}d^{3}z_{2}\partial_{z_{1}}^{\mu}\partial_{z_{2}}^{\nu}h(z_{1}-z_{2})S(x-z_{1})\gamma^{\mu}S(z_{1}-z_{2})\gamma^{\nu}S(z_{2}-y)\,,\label{eq:A.2}\end{equation}
 \begin{equation}
-i e^2 \alpha\int d^{3}z_{1}d^{3}z_{2}{\cal J}^{\nu}(z_{2})\partial_{z_{1}}^{\mu}\partial_{z_{2}}^{\nu}h(z_{1}-z_{2})\; S(x-z_{1})\gamma^{\mu}S(z_{1}-y)\,,\label{eq:A.3}\end{equation}
\begin{equation}
-\frac{e^2 \alpha}{2}\int d^{3}z_{1}d^{3}z_{2}{\cal J}^{\mu}(z_{1}){\cal J}^{\nu}(z_{2})\partial_{z_{1}}^{\mu}\partial_{z_{2}}^{\nu}h(z_{1}-z_{2})S(x-y)\,.\label{eq:A.4}\end{equation}
Using eq.\ (\ref{eq:2.10}), the expressions in 
eqs.\ (\ref{eq:A.4}) and (\ref{eq:A.3}) are, respectively, \begin{eqnarray}
 &  & -e^{2}\alpha S(x-y)\ob{h(0)-h(x-y)}\,,\label{eq:A.5}\\
 &  & -i e^{2}\alpha\int d^{3}z_{1}\partial_{z_{1}}^{\mu}\ob{S(x-z_{1})\gamma^{\mu}S(z_{1}-y)}\ob{h(x-z_{1})-h(y-z_{1})}\,.\label{eq:A.6}\end{eqnarray}
(We use integration by parts wherever it is useful.)
 Since $i\rlap/\partial_{x}S(x-y)=\delta^{(3)}(x-y)
 =iS(x-y){\stackrel\leftarrow{\rlap/\partial_{x}}}$,
eq.\ (\ref{eq:A.6})
can be written as \begin{equation}
2e^{2}\alpha S(x-y)\ob{h(0)-h(x-y)},\label{eq:A.7}\end{equation}
 whereas eq.\ (\ref{eq:A.2}) becomes \begin{eqnarray}
 -ie^2 \alpha \Bigg(\int d^{3}z_{1}
(\partial_{z_{1}}^{\mu}h(z_{1}-z_{2}))\Big|_{z_{2}=z_{1}}
S(x-z_{1})\gamma^{\mu}S(z_{1}-y)\nonumber\\
 -\int d^{3}z_{1}
(\partial_{z_{1}}^{\mu}h(y-z_{1}))
S(x-z_{1})\gamma^{\mu}S(z_{1}-y)
\Bigg)\,.\label{eq:A.8}\end{eqnarray}
 Now $\partial_{z_{1}}^{\mu}h(z_{1}-z_{2})$ is odd in $z_{1}-z_{2}$
and hence vanishes at $z_{1}=z_{2}$. Therefore we are left with the
second term which becomes \begin{equation}
-e^{2}\alpha S(x-y)\ob{h(0)-h(x-y)}\,.\label{eq:A.9}\end{equation}
 We see that the sum of eq.\ (\ref{eq:A.5}), eq.\ (\ref{eq:A.7}) and
eq.\ (\ref{eq:A.9}) vanishes, showing independence from the gauge parameter.

From the expressions for the leading order contributions, it 
is also easy to see that in the case of the isotropic dressing,
the contribution is simply from the usual fermion self energy in the Landau gauge.
Since ${\cal J}^\mu$ is a pure gradient in this case, integration by parts in 
both the contributions eq.\ (\ref{eq:A.3}) and eq.\ (\ref{eq:A.4})
of the photon cloud 
(with $\Delta^{\mu\nu}(x,y)$ instead of the expression (\ref{eq:A.1}))
shows that the dependence on the photon propagator is 
of the form
$\partial_{z_2}^{\nu}\Delta^{\mu\nu}(z_1,z_2)$.
This vanishes in the Landau gauge.

We now demonstrate {\it the problems with the Wilson line and the Dirac dressing}. For this
it is convenient to use the momentum-space representations as given in 
eqs.\ (\ref{eq:2.11})-(\ref{eq:2.13}). First, we give some details on deriving these
formulas. These are obtained by expressing
$J_\mu$ and $A_\mu$ in eqs.\ (\ref{eq:2.9}) and (\ref{eq:2.95}) in terms of their Fourier components: 
\bea
{\cal S}_{\alpha\beta}(x,y)=<\psi_{\alpha}(x)\bar{\psi}_{\beta}(y)
\exp\ob{-ie\int\frac{d^{3}l}{(2\pi)^{3}}(e^{il.x}-e^{il.y}){\tilde J}_\mu(l)\tilde{A}_{\mu}(l)}>\,.
\eea
For the isotropic case, eqs.\ (\ref{eq:2.4}) and (\ref{eq:2.55}) lead to ${\tilde J}_\mu(l)=il_\mu/l^2$.
For the Dirac dressing of eq.\ (\ref{eq:2.65}), ${\tilde J}_i(l)$ is independent of $l_3$ and
is the two-dimensional Fourier transform of $j_i$. Because of isotropy (in two dimensions),
${\tilde J}_i(l)=il_i/{\vec l}^2$. To arrive at the Wilson line case of eq.\ (\ref{eq:2.13}),
it is convenient to express $A_3(z)$ in eq.\ (\ref{eq:wil}) in terms of ${\tilde A}_3(l)$, 
and perform the integration over $z_3$.

To obtain the contribution of the photon cloud self-energy,
we expand to $O(e^2)$ the exponential of the appropriate expression from
eqs.\ (\ref{eq:2.11})-(\ref{eq:2.13}),
and use $< {\tilde A}_\mu(l) {\tilde A}_\nu(l')>
=(2\pi)^3 \delta^{(3)}(l+l')\Delta_{\mu\nu}(l)$, with
\bea
\Delta_{\mu\nu}(l)=\frac{\delta_{\mu\nu}-(1-\alpha)l_\mu l_\nu/l^2}{(l^2+\mu l)(1+l^2/\Lambda^2)}\,,
\eea 
the photon propagator in the $1/N$ expansion ($\mu=Ne^2/8$). 
Here we have used a regularization using a higher derivative for
the photon propagator \cite{lee}.
[For this, the following terms are to be added to the Lagrangian:
\bea
\frac{1}{\Lambda^2} \Bigg[\frac{1}{4}
(\partial_\sigma F_{\mu\nu})_x 
\Big(1+\frac{\mu}{\sqrt{-\partial^2}}\Big)_{xy} 
(\partial_\sigma F_{\mu\nu})_y 
+\frac{1}{2\alpha}
(\partial_\sigma \partial_\mu A_\mu)_x
\Big(1+\frac{\mu}{\sqrt{-\partial^2}}\Big)_{xy} 
(\partial_\sigma \partial_\nu A_\nu)_y\Bigg]
\label{Ladd}\eea
(an integration over repeated spacetime index is implied). Note that the first term in 
eq.\ (\ref{Ladd}) is gauge-invariant, while the second term modifies the gauge-fixing term.]

For the Wilson line, which we take to extend from $(0,0,x_3)$ to $(0,0,y_3)$, the
contribution of the photon cloud self-energy from eq.\ (\ref{eq:2.13}) to this order is then
\bea
S(x-y)\exp\left[-\frac{e^2}{2}\int\frac{d^{3}l}{(2\pi)^{3}}\frac{\ob{e^{il_{3}x_3}-e^{il_{3}y_3}}\ob{e^{-il_{3}x_3}-e^{-il_{3}y_3}}}{(l_{3})^{2}(l^{2}+\mu l)(1+l^2/\Lambda^2)}\right]\,.\label{wilson}\eea
We have chosen the Feynman gauge to simplify our calculations (also, the part of $\Delta_{\mu\nu}(l)$
proportional to $l_\mu l_\nu$ cancels out between the three contributions).
Using spherical coordinates, the exponent in eq.\ (\ref{wilson}) becomes  
\bea
-\frac{e^2}{\pi^2}
\int_{0}^{\infty}dl\frac{1}{(l^2+\mu l)(1+l^2/\Lambda^2)}\int_{0}^{1}d\eta\frac{4\sin^{2}(lr\eta/2)}{\eta^{2}}\eea
 where $\eta=\cos\theta$ and $r=|x_3-y_3|$ is the length of
the Wilson line. In terms of the new variables $\rho=lr$ and $\xi=lr\eta$, this becomes
\bea
-\frac{e^2r}{\pi^2}
\int_{0}^{\infty}d\rho\frac{1}{(\rho+\mu r)(1+\frac{\rho^2}{\Lambda^2 r^2})}\int_{0}^{\rho}d\xi\frac{\sin^{2}(\xi/2)}{\xi^{2}}\eea
 Now the integral over $\xi$ is bounded as $\rho\rightarrow\infty$.
Therefore with our regularization for the photon propagator, the
double integral exists. Call this double integral $I$ (without the $-e^2r/\pi^2$ in front).
For $\Lambda r\rightarrow\infty$, $I$ diverges logarithmically; so we expect $I\sim \ln (\Lambda r)$
for large $r$. To confirm this behaviour, we let $(\Lambda r)^{-1}\equiv\epsilon$
and find $-[\epsilon (dI/d\epsilon)]|_{\epsilon=0}$. In this process, after obtaining
$-\epsilon (dI/d\epsilon)$, it is convenient to put $\rho=\sigma/\epsilon$ and then take $\epsilon\rightarrow 0$.
This gives the coefficient of $\ln(\Lambda r)$ in $I$ as
\bea
2\int_{0}^{\infty}d\sigma\frac{\sigma}{(1+\sigma^{2})^2}\int_{0}^{\infty}d\xi\frac{\sin^{2}(\xi/2)}{\xi^{2}}\,,
\eea
 which is finite and non-zero.

Thus there is an overall behaviour $e^{-c r}$ for large $r$
when the stringy ansatz is used for the gauge-invariant dressing.
Such a linear term has been seen in earlier calculations \cite{ye} in the covariant
gauge with a $1/(\mu l)$ photon propagator and a momentum cut-off. But it has been regarded as a spurious
effect due to the regularization not being gauge invariant. In our
calculation, we have used the full $1/(l^2+\mu l)$ propagator with a gauge invariant regularization, and the
linear term survives.

If one chooses the temporal gauge $A^{3}=0$, it appears that the string
can be gauged away. This, of course, does not contradict the fact that for the stringy ansatz
$J^\mu$ has a curl part coupling to the gauge-invariant degrees of the photon.
The gradient part of $J^\mu$ couples to $\partial^\mu A^\mu$, and in the temporal
gauge $\partial^\mu A^\mu$ gets tuned so that this part of the dressing exactly cancels
the gauge-invariant part. 
The temporal gauge is beset with ambiguities. Otherwise,
a careful calculation should give back the term linear in $r$ 
from the fermion self-energy graph itself.

We now consider the Dirac dressing given in eq.\ (\ref{eq:2.12}).  Consider the case $x=(0,0,0)$
and $y=(0,0,r)$, i.e., the propagator at equal spatial coordinates.
The dressing is in the spatial planes at $x_{3}=0$ and $x_{3}=r$.
The contribution of the self-energy of the photon cloud is
\bea
S(x-y)\exp\left[-2e^2\int\frac{d^3l}{(2\pi)^3}\frac{1}
{\vec l^2(l^2+\mu l)(1+l^2/\Lambda^2)}
\sin^2\frac{l_3r}{2}
\right]\label{eq:41}
\eea
 where $\vec l=(l_{1},l_{2})$ is the spatial momentum. Holding $l_3$ fixed at a finite value,
we notice immediately
the logarithmic iR ($\vec l\rightarrow 0$) divergence $\int d^{2}\vec l/\vec l^{2}$. This is the
infinite self-energy of the two-dimensional Coulomb field in the spatial
planes $x_{3}=0$ and $x_{3}=r$ of infinite extent. Thus the Dirac
dressing does not give a finite propagator even for finite $r$. (If we consider
propagation in spatial direction, taking $x=(0,0,0)$ and $y=(r,0,0)$, 
we get a finite propagator. The reason is that the two-dimensional Coulomb
field from equal and opposite charges falls off rapidly to yield an
iR finite self energy.)

It is instructive to see that the contribution from the photon cloud self-energy is free of
divergence for the isotropic dressing. This contribution is:
\bea
S(x-y)\exp\left[-2e^2\alpha\int\frac{d^3l}{(2\pi)^3}\frac{1}
{l^2(l^2+\mu l)}
\sin^2\frac{l\cdot(x-y)}{2}
\right]\,.\label{eq:42}
\eea
The integral in the exponent is finite at both the ends $l\rightarrow \infty$ and $l\rightarrow 0$.
It can be shown \cite{shown}
that for $|x-y|\rightarrow\infty$, the exponent is $\ln(\mu|x-y|)$ upto a constant of proportionality, so that
exponentiation leads to a power law behaviour from this contribution.


\begin{thebibliography}{10}
\bibitem{not}
We follow the notation of Ref.\ \cite{mrs1}.
\bibitem{xi}
$\alpha$ is related to the gauge parameter $\xi$ used in Ref.\ \cite{mrs} by
$\alpha=1-\xi$.
\bibitem{pos}
$Z(p)\sim p^\eta$ with $\eta < 0$ would violate $0\le Z(0) <1$
 (see Ref.\ \cite{khv1}).
\bibitem{shown}
We can replace $\sin^2(l\cdot(x-y)/2)$ by $(1/2)(1-\exp(il\cdot(x-y)))$ in the integral
of eq.\ (\ref{eq:42}); this reduces it to the integral in $f(x-y)$ of Ref.\ \cite{mrs1}.
\end{thebibliography}
\end{document}